\documentclass[conference]{IEEEtran}
\usepackage{graphicx}

\usepackage{amsfonts}
\usepackage{amsthm}

\usepackage[normalem]{ulem}
\usepackage{color}
\usepackage{cite}
\usepackage{amsmath}
\usepackage{subfigure}
\usepackage{tabularx}
\usepackage{booktabs}
\usepackage{footnote}
\usepackage{algorithm}
\usepackage{algorithmic}
\usepackage{epstopdf}
\usepackage{array}
\usepackage{longtable}
\usepackage{tabu}
\usepackage{booktabs}
\usepackage{diagbox}
\usepackage{tabulary}
\usepackage{rotating}
\usepackage{multirow}
\usepackage{colortbl}
\usepackage{fancyhdr}
\usepackage{amssymb}
\usepackage{enumitem}
\usepackage{epstopdf}
\usepackage{stfloats}
\usepackage{soul}
\usepackage{mdwmath}
\usepackage{mdwtab}
\usepackage{mathtools}
\usepackage{caption}
\usepackage{rotating}
\renewcommand\thesubsection{\thesection.\Alph{subsection}}
\graphicspath{{C:/Users/admin/Documents/MATLAB/RS OBB/}}
\ifCLASSINFOpdf
\else
\fi
\usepackage{cite}

\begin{document}
\hyphenation{op-tical net-works semi-conduc-tor}
\makeatletter
\renewcommand\normalsize{%
	\@setfontsize\normalsize\@xpt\@xiipt
	\abovedisplayskip 4.5\p@ \@plus2\p@ \@minus5\p@
	\abovedisplayshortskip \z@ \@plus3\p@
	\belowdisplayshortskip 4.5\p@ \@plus3\p@ \@minus3\p@
	\belowdisplayskip \abovedisplayskip
	\let\@listi\@listI}
\normalsize
\renewcommand{\thefootnote}{}
%\title{Multigroup Multicast with a Rate-Splitting Beamforming in Multicarrier Systems}
\title{Rate-Splitting for Multigroup Multicast Beamforming in Multicarrier Systems}

%\author{\IEEEauthorblockN{Authors\\University of Surrey, Imperial College London, UK}
%	()@surrey.ac.uk}
%\author{\IEEEauthorblockN{Hongzhi~Chen\IEEEauthorrefmark{1}, De~Mi\IEEEauthorrefmark{1}, David~Vargas\IEEEauthorrefmark{2}, Zheng~Chu\IEEEauthorrefmark{1}, Bruno~Clerckx\IEEEauthorrefmark{3} and Pei~Xiao\IEEEauthorrefmark{1}\\5G Innovation Center, University of Surrey, UK}
%	Email@surrey.ac.uk}



%\author{\IEEEauthorblockN{Hongzhi~Chen\IEEEauthorrefmark{1}, De~Mi\IEEEauthorrefmark{1}, Zheng~Chu\IEEEauthorrefmark{1}, Pei~Xiao\IEEEauthorrefmark{1} and Bruno~Clerckx\IEEEauthorrefmark{2}}
%\IEEEauthorblockA{\IEEEauthorrefmark{1}Institute for Communication Systems, University of Surrey, United Kingdom}
%\IEEEauthorblockA{\IEEEauthorrefmark{2}Department of Electrical and Electronic Engineering, Imperial College London, United Kingdom}
%Email:\{hongzhi.chen, d.mi, zheng.chu, p.xiao\}@surrey.ac.uk, b.clerckx@imperial.ac.uk}

\author{\IEEEauthorblockN{Hongzhi~Chen, De~Mi, Zheng~Chu, Pei~Xiao and Rahim~Tafazolli}
	\IEEEauthorblockA{Institute for Communication Systems, University of Surrey, United Kingdom}
	%\IEEEauthorblockA{\IEEEauthorrefmark{3}Department of Electrical and Electronic Engineering, Imperial College London, United Kingdom}
	Email:\{hongzhi.chen, d.mi, zheng.chu, p.xiao, r.tafazolli\}@surrey.ac.uk}


\maketitle
\begin{abstract}
	
	In this paper, we consider multigroup multicast transmissions with different types of service messages in an overloaded multicarrier system, where the number of transmitter antennas is insufficient to mitigate all inter-group interference. We show that employing a rate-splitting based multiuser beamforming approach enables a simultaneous delivery of the multiple service messages over the same time-frequency resources in a non-orthogonal fashion. Such an approach, taking into account transmission power constraints which are inevitable in practice, outperforms classic beamforming methods as well as current standardized multicast technologies, in terms of both spectrum efficiency and the flexibility of radio resource allocation.
	%This paper considers an overloaded multicarrier system to support multigroup multicast transmissions with different types of service contents/objects, in the presence of the inter-group interference and transmission power constraint, which inevitable in practice. We show that employing a rate-splitting based multiuser beamforming approach enables a simultaneous delivery of the multiple service contents/objects over the same time-frequency resources in a non-orthogonal fashion, thus in a spectrum-efficient way, as well as provides a flexible resource allocation with the aid of an effective subcarrier selection strategy. %, offering an improved energy efficiency.
	
\end{abstract}

%\begin{IEEEkeywords}
%	Multigroup multicast, multicarrier, rate-splitting, multi-service.
%\end{IEEEkeywords}

% \fi
\IEEEpeerreviewmaketitle

{\scriptsize \footnote{We would like to acknowledge the support of the University of Surrey 5GIC (www.surrey.ac.uk/5gic) members for this work. This work was also supported in part by the European Commission under the 5GPPP project 5G-Xcast (H2020-ICT-2016-2 call, grant number 761498). The views expressed in this contribution are those of the authors and do not necessarily represent the project.}}

\section{Introduction}
Point-to-multipoint communication capabilities have been kept enhancing in the legacy 4G LTE (4th Generation Long-Term Evolution) since the adoption of eMBMS (evolved Multimedia Broadcast Multicast Service) in 3GPP (3rd Generation Partnership Project) Release 9. One representative transmission mode, multicast, aims to provide an identical content to a group of interested users or distinct contents to multiple groups of users simultaneously, via the existing radio network infrastructure of 4G LTE with minor changes. Due to its efficient use of the available spectrum resources in delivering the common content, multicast has been identified as an essential candidate in the development of 5G (5th Generation) wireless communications towards vertical use cases, e.g., multicast public warning alert and object-based broadcasting (OBB) \cite{D31}. In particular, OBB enables a more interactive media experience for subscribed users, by turning traditional multicast programs into different types of service messages, named as elements, which allows the users to reassemble programs based on local factors, for example, the device or environment \cite{BBCOBB}.%e.g., background contents and personalized objects including metadata, which allows the users to reassemble programs based on local factors, for example, the device or environment \cite{BBCOBB}.

Despite the fact that the current multicast technologies are considerably different compared with the original versions in Release 9, they still carry a long legacy thanks to the backwards-compatible design principle of 4G LTE. For example, the current multicast systems in the eMBMS use OFDM (Orthogonal Frequency Division Multiplexing) as the carrier waveform and Time Division Multiplexing (TDM) to separate different services, which, however, can cause the inefficient use of time-frequency resources in the emerging 5G use cases like OBB. More specifically, the aforementioned OBB elements in one program can be allocated in different subcarriers, but different programs must be transmitted in different time intervals in the current TDM-based multicast systems. %the aforementioned media objects and metadata in OBB are generally treated as multicast and unicast contents respectively, and as a result. it is required to allocate the aforementioned %background contents, personalized ...different types of OBB service messages in different time intervals in the current TDM-based multicast systems. On the other hand, in order to mitigate the inter-group interference in the case of the multigroup multicast, 
In the case of the multigroup multicast, in order to realize the simultaneous delivery of the different multicast programs for the corresponding groups of users in the same time and frequency resources, classic beamforming methods require a sufficient number of transmitter antennas (offering a sufficient number of spatial degree of freedom) to neutralize the inter-group interference. Nevertheless, such requirement can be difficult in practical systems where it is inevitable to serve a relatively large number of groups/users, and can be even more challenging for the multicast technologies in the current eMBMS standardized with a restricted number of transmitter antennas \cite{D31}. %, e.g., up to 2 \cite{D31}. 
Therefore, it is of great interest to design an efficient multiuser beamforming that simultaneously delivers the different types of service messages by using same time-frequency resources, for the multigroup multicast scenarios where the system relies on the multicarrier transmission like OFDM and is highly likely overloaded.

%To this end, we propose a Rate-Splitting (RS) multigroup multicast beamforming approach in an overloaded multicarrier system, where the RS strategy 
Recently, Rate-Splitting (RS) based multicast beamforming strategy has been considered \cite{RSMMFMM}\cite{RSOLsystem} as an efficient way to deliver the different types of service messages to multiple groups/users in the presence of the inter-group interference. In the RS strategy, each service message intended for one group, e.g., the OBB element from one multicast program, can be split into a broadcast part and a multicast part. All broadcast parts are superposed to form a super transmitted message and broadcasted to all users, while multicast parts are delivered based on the conventional beamforming methods. The main contribution of this work is that we generalize the optimization of RS precoders in \cite{RSMMFMM}\cite{RSOLsystem} to overloaded multicarrier scenarios by introducing a total transmission power constraint across all subcarriers. Such extension fills the research gap among the existing literature on employing the RS multigroup multicast beamforming approach in multicarrier systems.  %Motivated by the strong relationship between the RS strategy and design philosophy of the OBB, we consider the RS multigroup multicast beamforming approach in an overloaded multicarrier system to support the OBB-like use cases. 
Simulation results show that applying RS in the considered system not only enables the support of multi-service with a significantly improved spectrum efficiency, but also provides an increased flexibility to allocate radio resources and the restricted transmission power, compared with the classic beamforming methods as well as the standardized TDM-based multicast technologies. 

\section{System Model and Problem Formulation}
We consider a multicarrier downlink system comprising a single base station equipped with ${N}_t$ transmit antennas and $K$ single-antenna users (${N}_t < K$). These $K$ users are grouped into $M$ multicast groups, i.e., \{$\mathfrak{g}_1,\mathfrak{g}_2,...,\mathfrak{g}_M$\}, where $\mathfrak{g}_m$ is the set of users belonging to the $m$-th group. Users in the same group are interested in the same program so there are total $M$ programs i.e., OBB programs. It is assumed that one user can not belong to two groups. The number of available subcarriers is $N$. Each OBB program can be de-constructed into $N$ elements, and the $N$ available subcarriers are responsible for carrying the $N$ de-constructed OBB elements respectively. For traditional TDM based OBB program transmission, each program occupies one time slot and is transmitted to one specific multicast group, therefore totally $M$ time slots are needed. Instead of the conventional TDM-based methods for the OBB, we are particularly interested in a transmission mode which enables the simultaneous transmission of these $M$ distinct programs to specific groups within the same frequency resource, and we define this transmission mode as multicarrier multigroup multicasting OBB. More specifically, on each subcarrier, corresponding element from each program that are originally occupied $M$ time slots are now precoded and transmitted to all $K$ users in a single time slot, same applied to the rest $N-1$ subcarriers. By denoting the size of the $m$-th group as $G_{m}=|\mathfrak{g}_m|$, on each subcarrier:
\begin{equation}
G_{1}=...=G_{m}=...=G_{M}={K}/{M}
\end{equation} 
which means equal user grouping.   %but the user grouping may be different on each subcarrier. 
On each subcarrier (we use $n$ to denote subcarrier index), transmitter intends to send the $n$-th part of the original group messages (from now on, we refer the precoded elements transmitted on the $n$-th subcarrier and required by $m$-th group as the $n$-th part of the original group message, denoted as: $W_{M,n}$) $\{W_{1,n},...,W_{M,n}\}$ to $\{\mathfrak{g}_{1},...,\mathfrak{g}_{M}\}$, respectively. These $M$ messages are first encoded into independent symbol streams as $\mathbf{s}_{m,n}=\{s_{1,n},...,s_{M,n}\}$, which are then beamformed as:
\begin{equation}
\mathbf{x}_n = \sum_{m=1}^{M}\mathbf{p}_{m,n}\mathbf{s}_{m,n}
\end{equation}
where, we assume $\mathbb{E}\{\mathbf{s}_{m,n} \mathbf{s}_{m,n}^H\}=\mathbf{I}$, $\mathbf{x}_n\in\mathbb{C}^{N_t}$ represents the signal vector transmitted in a given channel. $\mathbf{p}_{m,n}\in\mathbb{C}^{N_t} $ represents the beamforming matrix for the $m$-th group.
The signal received by the $k$-th user is given as:
\begin{equation} \label{eq:ykn}
\mathbf{y}_{k,n}=\mathbf{h}^T_{k,n}\mathbf{x}_n+\mathbf{n}_{k,n }.
\end{equation}
The vector $\mathbf{n}_{k,n}\sim\mathcal{CN}(0,\sigma^2_{k,n})$ represents the received Additive White Gaussian Noise (AWGN) on the corresponding frequency band. Without loss of generality we assume %$\sigma^2_{1,1}=...=\sigma^2_{K,1}=...=\sigma^2_{K,N}$
$\sigma^2_{k,n}=\sigma^2, \forall k\in K,  \forall n\in N$. In \eqref{eq:ykn}, $\mathbf{h}_{k,n}\in\mathbb{C}^{N_t}$ is the channel gain between the $k$-th user and the transmitter on $n$-th subcarrier. We further assume that transmitter has the perfect channel state information, and the channel remains constant in one transmission period.
The total power constraint is
\begin{equation}
\sum_{n=1}^{N}\mathbb{E}\{\mathbf{x}_n \mathbf{x}_n^H\}=\sum_{n=1}^{N}P_n\leq P
\end{equation}
where $P_n$ represents the power allocated on the $n$-th subcarrier and $P_n\geq0$. 
Thus, the Signal to Interference plus Noise ratio (SINR) experienced by the $k$-th user ($k \in \mathfrak{g}_m$) on the $n$-th subcarrier, is given by:
\begin{equation}
\text{SINR}_{k,n}=\frac{|\mathbf{h}^T_{k,n}\mathbf{p}_{m,n}|^2}{\sum_{j\neq m}|\mathbf{h}^T_{k,n}\mathbf{p}_{j,n}|^2+\sigma^2_{k,n}}.
\end{equation}
The achievable rate of the $k$-th user is ${R}_{k,n}=\log_2(1+\text{SINR}_{k,n})$. To guarantee that all users in the $m$-th group can successfully decode the corresponding message, the transmission rate of the $n$-th part of group message should equal to the lowest achievable rate, given as:
\begin{equation}
R_{m,n} = \min_{k \in \mathfrak{g}_m}{R}_{k,n}.
\end{equation}
The beamforming matrix optimization on each subcarrier is based on a Max-Min Fair (MMF) principle, aiming to maximize the minimum transmission rate among all multicast groups, which benefits the multicast data rate, i.e., the multicast rate is decided by the lowest user data rate. So the subcarrier MMF rate among all $M$ groups is given by: $R_n = \min_{m \in M}{R}_{m,n}$, therefore the sum MMF-rate for the multicarrier multigroup multicast transmission is given by $R_{sMr} = \sum_{n=1}^{N}R_n$. And we formulate the sum MMF-rate optimization problem with classic beamforming in the next subsection.
\subsection{Classic Beamforming}
MMF optimization achieves balanced group rate on each subcarrier, subject to a total power constraint across all subcarriers. The sum MMF-rate can be formulated as:%We select two typical messages from the OBB content, i.e. background and objects, and then 
\begin{equation} \label{eq:RP} 
{R}_{sMr}(P): 
\begin{cases}
\max \limits_{\mathbf{P}} \quad  \sum_{n=1}^{N}\min\limits_{m \in M}\min\limits_{k \in \mathfrak{g}_m}{R}_{k,n} \\
\begin{aligned}
s.t. \quad
%  &{R}_{k,n_b} \!\geq\! {R}^b, \quad \forall k \in K\\
%  &{R}_{k,n_o} \!\geq\! {R}^o, \quad \forall k \in K \\  
  &\sum_{n=1}^{N}\sum_{m=1}^{M}||\mathbf{p}_{m,n}||^2 \leq P,\\ 
\end{aligned}
\end{cases}
\end{equation}where $\mathbf{P} \triangleq (\mathbf{p}_{1,1},...,\mathbf{p}_{m,n},...\mathbf{p}_{M,N})$ is set of precoding matrix. %More generally, it can be any elements that construct the OBB content and required by more than one users.
The inner minimization in \eqref{eq:RP} denotes the multicast rate for the $m$-th group while the outer minimization accounts for the fairness across all $M$ groups, and the summation accounts the sum MMF-rate across all $N$ subcarriers.  
Problem \eqref{eq:RP} can be treated as a combination of $N$ parallel MMF optimization problems and can be solved by SDR method \cite{SDR} yet resulting in the saturated sum-MMF-rate.
\section{Rate-Splitting Beamforming for Multicarrier Multigroup Multicasting}
\subsection{Rate-Splitting Scheme}
On each subcarrier, each group message is split into broadcast and multicast parts, i.e., $W_{m,n} \Rightarrow\{W_{m0,n},W_{m1,n}\}$, respectively, before precoding. 
%split $m$-th group message  which also split the group data rate $R_{m,n}$ into $R_{m0,n}$ and $R_{m1,n}$.
Then all broadcast parts form one super message to broadcast, namely broadcast message, i.e., $W_{bc,n} = \{W_{10,n},W_{20,n},...,W_{M0,n}\}$, which is received and decoded by all users. Denote the encoded broadcast symbol by $\mathbf{s}_{bc,n}$, the transmit signal with linear precoding can be expressed as:
\begin{equation}
\mathbf{x}_{n} =\mathbf{p}_{bc,n}\mathbf{s}_{bc,n} + \sum_{m=1}^{M}\mathbf{p}_{m,n}\mathbf{s}_{m,n}
\end{equation}
where $\mathbf{p}_{bc,n}$, $\mathbf{s}_{m,n}$ is the precoder for broadcast message and encoded multicast symbols respectively. 
%The signal power received by the $k$-th user can be written as:
%\begin{equation}
%\text{T}_{k,n}\!\!=\!|\mathbf{h}^T_{k,n}\mathbf{p}_{bc,n}|^2+\!\sum_{m=1}^{M}|\mathbf{h}^T_{k,n}\mathbf{p}_{m,n}|^2+\sigma^2_{k,n}
%\end{equation}
The total power constraint among all subcarrier is given by:
\begin{equation} {\label{eq:rspc}}
 \sum_{n=1}^{N}(||\mathbf{p}_{bc,n}||^2+\sum_{m=1}^{M}||\mathbf{p}_{m,n}||^2)\leq P.
\end{equation}
At the receiver side, the $k$-th user retrieves the corresponding multicast message in a successive interference cancellation (SIC) manner. The broadcast part is first decoded by treating the remaining multicast part as noise. After successfully decoding and removing the broadcast part, the SIC receiver then decodes the multicast part. RS scheme is degraded to non-RS algorithm when the rate of the broadcast part is zero. The achievable broadcast and multicast rate for the $k$-th user in $m$-th group can be expressed as:
{\small\begin{equation}
\begin{split}
&{R}_{bc,k,n}=\log_2(1+\frac{|\mathbf{h}^T_{k,n}\mathbf{p}_{bc,n}|^2}{\sum_{m=1}^{M}|\mathbf{h}^T_{k,n}\mathbf{p}_{m,n}|^2+\sigma^2_{k,n}})\\
&{R}_{k,n}=\log_2(1+\frac{|\mathbf{h}^T_{k,n}\mathbf{p}_{m,n}|^2}{\sum_{j \neq m}|\mathbf{h}^T_{k,n}\mathbf{p}_{j,n}|^2+\sigma^2_{k,n}}).
\end{split}
\end{equation}}
The rate of broadcast message, to guarantee the successful decoding of which, should equal to the minimum achievable rate among all users, i.e.,
${R}_{bc,n} = \min\limits_{\forall k \in K}{R}_{bc,k,n}$,
where ${R}_{bc,n} = \sum_{m=1}^{M}{C}_{m,n} $ and ${C}_{m,n}$ is the broadcast part extracted from the original group message and is intended to transmit to group $m$.
Thus, the RS based achievable rate for the $m$-th group on the $n$-th subcarrier is
\begin{equation}
{R}_{m,n} = {C}_{m,n}  + \min_{m \in M}\min_{k \in \mathfrak{g}_m}{R}_{k,n}.
\end{equation}
%However, the sum-MMF-rate for all subcarriers can still be expressed as (8) since we still need subcarrier allocation to optimize the sum-MMF rate.
\subsection{Sum MMF-Rate with Rate-Splitting}
The RS-based sum MMF-rate optimization problem is formulated in \eqref{eq:SRRS}, including additional constraints related to the broadcast rate and its distribution between different groups. 
{\small\begin{equation}  \label{eq:SRRS}
\text{R}_{sMr}^{RS}(P)\!\!:\!\! 
	\begin{cases}
	\max \limits_{\mathbf{C},\mathbf{P}_{RS}} 
	  \sum_{n=1}^{N} \min\limits_{m \in M} {R}_{m,n}\\ 
	\begin{aligned}
	s.t.\quad
	&{R}_{m,n} = {C}_{m,n}  + \min_{m \in M}\min_{k \in \mathfrak{g}_m}{R}_{k,n}.\\
	&{C}_{m,n} \geq 0, \forall m\in M, \forall n\in N\\
	&{R}_{bc,k,n} \geq \sum \limits_{m=1}^{M} \! \!{C}_{m,n}, \forall k\in K,\forall n\in N\\
	&\sum_{n=1}^{N}(||\mathbf{p}_{bc,n}||^2+\sum_{m=1}^{M}||\mathbf{p}_{m,n}||^2)\leq P
	\end{aligned}
	\end{cases}
	\end{equation}}where $\mathbf{P}_{RS} \triangleq (\mathbf{p}_{bc,1},\mathbf{p}_{1,1},...,\mathbf{p}_{M,n},...,\mathbf{p}_{bc,N},,...,\mathbf{p}_{M,N})$ is set of precoding matrix,  $\mathbf{C} \triangleq (\mathbf{C}_{1,1},...,\mathbf{C}_{m,n},...,\mathbf{C}_{M,N})$. The second and third set of constraints guarantee the non-negative valued broadcast message can be decoded by all $K$ users.  
\subsection{WMMSE based RS precoder and equalizer optimization}
Problem \eqref{eq:SRRS} involves a sum of two rate components on each subcarrier which cannot be solved by SDR method. Research shows that a WMMSE approach can efficiently solve problems with non-convex coupled summation of multiple rates \cite{RSMMFMM}\cite{RSOLsystem}\cite{MMSE2}, by building a connection of Weighted MSE and the corresponding rate. The weighted MSE for broadcast and multicast message are given by:
\begin{equation}
\begin{split}
&\text{W}_{\text{MSE}_{bc,k,n}} = v_{bc,k,n}\text{MSE}_{bc,k,n}-\log_2(v_{bc,k,n})\\
&\text{W}_{\text{MSE}_{k,n}} = v_{k,n}\text{MSE}_{k,n}-\log_2(v_{k,n})\\
\end{split}
\end{equation}  
where $v_{bc,k,n}$ and $v_{k,n}$ are positive weight variables. $\text{MSE}_{bc,k,n}$ and $\text{MSE}_{k,n}$ represents the corresponding Mean Square Errors (\text{MSE}s) for broadcast and multicast signals. To obtain the optimal weight \text{MSE} value, we calculate the partial derivative of the weighted \text{MSE} to the weight variable and the \text{MSE} separately, results in: 
$v_{bc,k,n}=(\ln 2*\text{MSE}_{bc,k,n})^{-1}$ and 
$v_{k,n}=({\ln 2*\text{MSE}_{k,n}})^{-1}$. 
Denoting the estimated broadcast and multicast signals received by the $k$-th user as $\widehat{\mathbf{s}}_{bc,k,n}$ and $\widehat{\mathbf{s}}_{k,n}$ respectively. 
By applying equalizers associated with the SIC receiver process, the estimated broadcast and multicast signal can be expressed as:
\begin{equation}
\begin{split}
&\widehat{\mathbf{s}}_{bc,k,n} = {g}_{bc,k,n}\mathbf{y}_{k,n} \\ &\widehat{\mathbf{s}}_{k,n} = {g}_{k,n}(\mathbf{y}_{k,n}-{g}_{bc,k,n}\mathbf{y}_{k,n})
\end{split}
\end{equation}
And we have:
{\small \begin{equation}
	\begin{split}
&\text{MSE}_{bc,k,n} = |{g}_{bc,k,n}|^2\text{T}_{k,n}-2\mathbb{R}\{{g}_{bc,k,n}\mathbf{h}^T_{k,n}\mathbf{p}_{bc,n}\}+1\\
&\text{MSE}_{k,n} = |{g}_{k,n}|^2\text{E}_{k,n}-2\mathbb{R}\{{g}_{k,n}\mathbf{h}^T_{k,n}\mathbf{p}_{m,n}\}+1\\
\end{split}
\end{equation}}where $\text{T}_{k,n}$ is the total received signal power by $k$-th user, and $\text{E}_{k,n}$ is the interference plus noise power when user decoding the broadcast signal. And the corresponding equalizers can be obtained as:
${g}_{bc,k,n}={\mathbf{h}^T_{k,n}\mathbf{p}_{bc,n}}/{\text{T}_{k,n}},
{g}_{k,n}={\mathbf{h}^T_{k,n}\mathbf{p}_{m,n}}/{\text{E}_{k,n}}$, which leads to: $\text{MSE}_{bc,k,n} =\frac{\text{E}_{k,n}}{\text{T}_{k,n}},
\text{MSE}_{k,n} = \frac{\text{Q}_{k,n}}{\text{E}_{k,n}}$
, where $\text{Q}_{k,n}$ represents the interference plus noise when user decoding the specific multicast signal. Re-writing the $\text{SINR}$ in (12):
\begin{equation}
\begin{split}
&\text{SINR}_{bc,k,n}\!=\!\frac{\text{T}_{k,n}}{\text{E}_{k,n}}\!-\!1,
\text{SINR}_{k,n}\!=\! \frac{\text{E}_{k,n}}{\text{Q}_{k,n}}\!-1\!\\
\end{split}
\end{equation}
which combines the \text{MSE}s and the corresponding rate by:
\begin{equation}
\begin{split}
&{R}_{bc,k,n}= \log_2(\frac{\text{T}_{k,n}}{\text{E}_{k,n}})=-\log_2(\text{MSE}_{bc,k,n})\\
&{R}_{k,n}= \log_2(\frac{\text{E}_{k,n}}{\text{Q}_{k,n}})=-\log_2(\text{MSE}_{k,n})\\
\end{split}
\end{equation} combining (17) and (19) yields:
\begin{equation} \label{eq:RWMSE}
\begin{split}
&\text{W}_{\text{MSE}_{bc,k,n}} = {G} - {R}_{bc,k,n}\\
&\text{W}_{\text{MSE}_{k,n}}   = {G} - {R}_{k,n}
\end{split}
\end{equation}where constant ${G}={1}/{\ln 2}+\log_2(\ln 2)$. Following the rate and Weighted MSE relationship in \eqref{eq:RWMSE}, \eqref{eq:SRRS} can be written as:
{\small\begin{equation} \label{eq:RS2} 
\text{R}_{sMr}^{RS}(P)\!\!:\!\! 
\begin{cases}
\max \limits_{{r}_{tot},\mathbf{C},\mathbf{P}_{RS},\mathbf{g},\mathbf{r},\mathbf{v}} \quad {r}_{tot}\\
\begin{aligned}
s.t.  \quad 
&\sum_{n=1}^{N} \min\limits_{m \in M} {R}_{m,n} \geq {r}_{tot}\\
&{C}_{m,n} \geq 0,\forall m\in M, \forall n\in N\\
&{C}_{m,n}+ {r}_{m,n} \geq  {R}_{m,n}\\ 
&{G}-\text{W}_{\text{MSE}_{bc,k,n}} \geq \sum_{m=1}^{M}{C}_{m,n}\\
&{G}-\text{W}_{\text{MSE}_{k,n}} \geq {r}_{m,n}\\
&\sum_{n=1}^{N}(||\mathbf{p}_{bc,n}||^2+\sum_{m=1}^{M}||\mathbf{p}_{m,n}||^2)\leq P
\end{aligned}
\end{cases}
\end{equation}}where $\mathbf{r} \triangleq ({r}_{1,1},...,{r}_{m,n},...,{r}_{M,N})$ are auxiliary variables, $\mathbf{v} \triangleq (v_{bc,k,n},v_{k,n})$ and  $\mathbf{g} \triangleq (g_{bc,k,n},g_{k,n})$. The proposed algorithm to solve the optimization problem (\ref{eq:RS2}) is presented in Algorithm 1.

\begin{algorithm}[H]
	\caption{Joint RS beamforming and equalizer optimization and sum MMF-rate optimization}
	\begin{algorithmic}[1]\label{alg2}
		{\small\STATE Results: $\sum_{n=1}^{N} \min\limits_{m \in M} {R}_{m,n}\geq {r}_{tot}$
		\STATE Initialize: $t=0$, $l=0$, $\mathbf{P}_{RS}$, ${r}_{tot}(t)=0$, ${R}_{n}(l)=0,\forall n \in N$;
		\STATE \textbf{repeat}
		\STATE \quad $t=t+1$;
		\STATE \quad\textbf{repeat}
		\STATE \quad \quad $l=l+1$;
		\STATE \quad \quad updating ${g}_{bc,k,n}$ and ${g}_{k,n}$, $\forall n \in N$; 
		\STATE \quad \quad updating $v_{bc,k,n}$ and $v_{k,n}$, $\forall n \in N$;
		\STATE \quad \quad solving optimization problem \eqref{eq:RS2}, which optimize $\mathbf{C},\mathbf{P},\mathbf{g},\mathbf{r},\mathbf{v}$
	    \STATE \quad \textbf{Until} $|{R}_{n}(l)-{R}_{n}(l-1)|\leq \varepsilon_1, \forall n \in N$ 
		\STATE \textbf{Until} $|{r}_{tot}(t)-{r}_{tot}(t-1)|\leq \varepsilon_2$}
	\end{algorithmic}
\end{algorithm} In algorithm 1, $\varepsilon_{1,2}$ and $(t,l)$ denote the error tolerance and the iteration indexes, respectively.
\section{Degree of Freedom Analysis}
\subsection{Define DoF}
On each subcarrier, the spatial degree of freedom (DoF) can be given by \cite{DOFformula} (Due to the orthogonality between subcarriers, the DOF for each subcarrier is the same. Therefore in this section, we conceal the subscripts of subcarriers.)
\begin{equation}
{d} = \lim\limits_{P\rightarrow\infty}\frac{\text{Ca}_{\sum}(P)}{\log_2(P)}
\end{equation}
where $\text{Ca}_{\sum}(P)$ is the sum capacity in case of a point of point communication at power level $P$. In other word, the degree of freedom denotes the number of interference-free channels. Therefore, for a single antenna user with corresponding beamforming matrix $\mathbf{P}(P)$, define ${R}_k(P)=\text{Ca}_{\sum}^{k}(P)$, thus
\begin{equation}
{d}_k = \lim\limits_{P\rightarrow\infty}\frac{{R}_k(P)}{\log_2(P)} \leq 1
\end{equation}
which stands for that, for a given set of achievable rate of K users, i.e., $\mathbf{R}=\{{R}_1(P),{R}_2(P),\ldots,{R}_K(P)\}$, we have the corresponding user DoF set as $\mathbf{d}=\{{d}_1,{d}_2,\ldots,{d}_K\}$, and the group DoF as $\mathbf{D}_g=\{{D}_1,{D}_2,\ldots,{D}_m\}$, where
$
{D}_m = \min \{{d}_k,{d}_{k+1},\cdots,{d}_{k+\frac{K}{M}}\},
 \{ {k},{k+1},\ldots,{k}+\frac{{K}}{{M}} \}  \in \mathfrak{g}_m
$. Thus the overall system DoF is given by:

\begin{equation}
\text{D}_{sys}=\min\{{D}_1,{D}_2,\ldots,{D}_m\}
\end{equation}
\subsection{DoF with Classic beamforming}
The design principle of the precoding/beamforming matrix is to null the corresponding interference. Therefore, the optimal, if applicable, beamforming matrix for the $m$-th group has to satisfied that
\begin{equation}
\begin{split}
&\mathbf{p}_m \perp \mathbf{H}_{\forall q,q \neq m}\Longrightarrow \mathbf{p}_m \in null(\mathbf{H}_{\bar m})
\end{split}
\end{equation}
where $\mathbf{H}_{\bar m}=\left[\mathbf{H}_1,\mathbf{H}_{2},\ldots,\mathbf{H}_{m-1},\mathbf{H}_{m+1},\ldots,\mathbf{H}_M \right]$ is a $N_t$-by-$(K-\frac{K}{M})$ matrix which has the null space dimension greater or equally to 1, the requirement on $N_t$ is:
$N_t-(K-\frac{K}{M}) \geq 1 \Longrightarrow N_t \geq 1+(K-\frac{K}{M})$.
Otherwise, all users may experience interference from all other groups. The degree of freedom can be summarised:
\begin{equation}
\text{D}_{sys}=
\begin{cases}
1, \quad M=1 \:\text{or}\: N_t \geq 1+(M-1)*(\frac{K}{M})\\
0, \quad N_t \leq 1+(M-1)*(\frac{K}{M})
\end{cases}
\end{equation}
where $M=1$ denotes an interference-free transmission and the second case mainly depends on the system load. This is the reason that under the overloaded situation, users may suffer very low rate growth w.r.t SNR growth, i.e., performance saturation, which will be shown in the simulation.
\subsection{DoF with RS}
The broadcast message has a useful portion to any specific user or group and can be successfully decoded by all users, which means it can contribute to the user's achievable rate. This explains the fact that how RS actually improves DoF bottleneck, more specifically:
\begin{itemize}
	\item The $M$ groups are categorised into \emph{degraded groups} and \emph{designated groups} i.e. $M = \left[M_{deg},M_{des}\right]$.
	\item Users in the $M_{deg}$ group act as degraded users who are willing to receive the message intended for other groups, which relaxes the requirement of the number of the transmit antennas to null the interference experienced by the users in the $M_{des}$ group, who can then enjoy the interference-free transmission.%Users in the group $M_{deg}$ are willing to receive the message intended for other groups, which enables  interference experienced by $M_{des}$ group of users.  
\end{itemize}

To analysis the DoF of RS, the total transmit power is divided into two portions $(P^{\beta}, P-P^{\beta})$, assigned for $M_{des}$ and $M_{deg}$ ($\beta$ is a power allocation factor). For $k \in M_{deg}$, the corresponding $\text{SINR}$ can be expressed as:
\begin{equation}
\text{SINR}_{k \in M_{deg}}=\frac{P-P^{\beta}}{P^{\beta}+1} \approx \frac{P-P^{\beta}}{P^{\beta}}\approx P^{1-\beta}
\end{equation}
the total DoF for all degraded groups is $\lim\limits_{P\rightarrow\infty}\frac{\log_2(P^{1-\beta})}{\log_2(P)}=({1-\beta})$, then each degraded group has $ \text{D}_{m \in M_{deg}}=\frac{1-\beta}{M_{deg}}$. After removing the broadcast message, each designated group has DoF: $\beta$. Hence, the overall system DoF is given by:
\begin{equation} \label{eq:dofsys} 
\text{D}_{sys}=\min\{\frac{1-\beta}{M_{deg}}, \beta\}
\end{equation}
where $\beta$ can be calculated by:
\begin{equation}
\frac{1-\beta}{M_{deg}}= \beta \Longleftrightarrow \beta=\frac{1}{1+M_{deg}}
\end{equation}
As a summary, for a given number of transmitter antennas, the highest achievable DoF depends on $\min\{M_{deg}\}$. Also note that $\text{D}_{sys} = \frac{1}{1+M_{deg}} \ge 0$ is promised, which reveals the fact that RS avoids the performance saturation from a DoF perspective. Also, by having $M_{deg} \leq (M-1) \Longrightarrow \text{D}_{sys}\geq \frac{1}{M}$, RS at least has the same performance as the basic power domain non-orthogonal multiple access. 
\section{Simulation Results}
The performance of the considered RS-based multicarrier beamforming strategies and the current eMBMS system with TDM, is fairly compared in this section, with the identical total power constraint, i.i.d. channels with entries drawn from $\mathcal{CN}(0,1)$. Results are averaged over 100 channel realizations. All the optimization problems that presented in convex form are solved using MATLAB based CVX toolbox \cite{CVX}.

\begin{figure}[t]
	\centering
	\includegraphics[angle=0,width=0.45\textwidth]{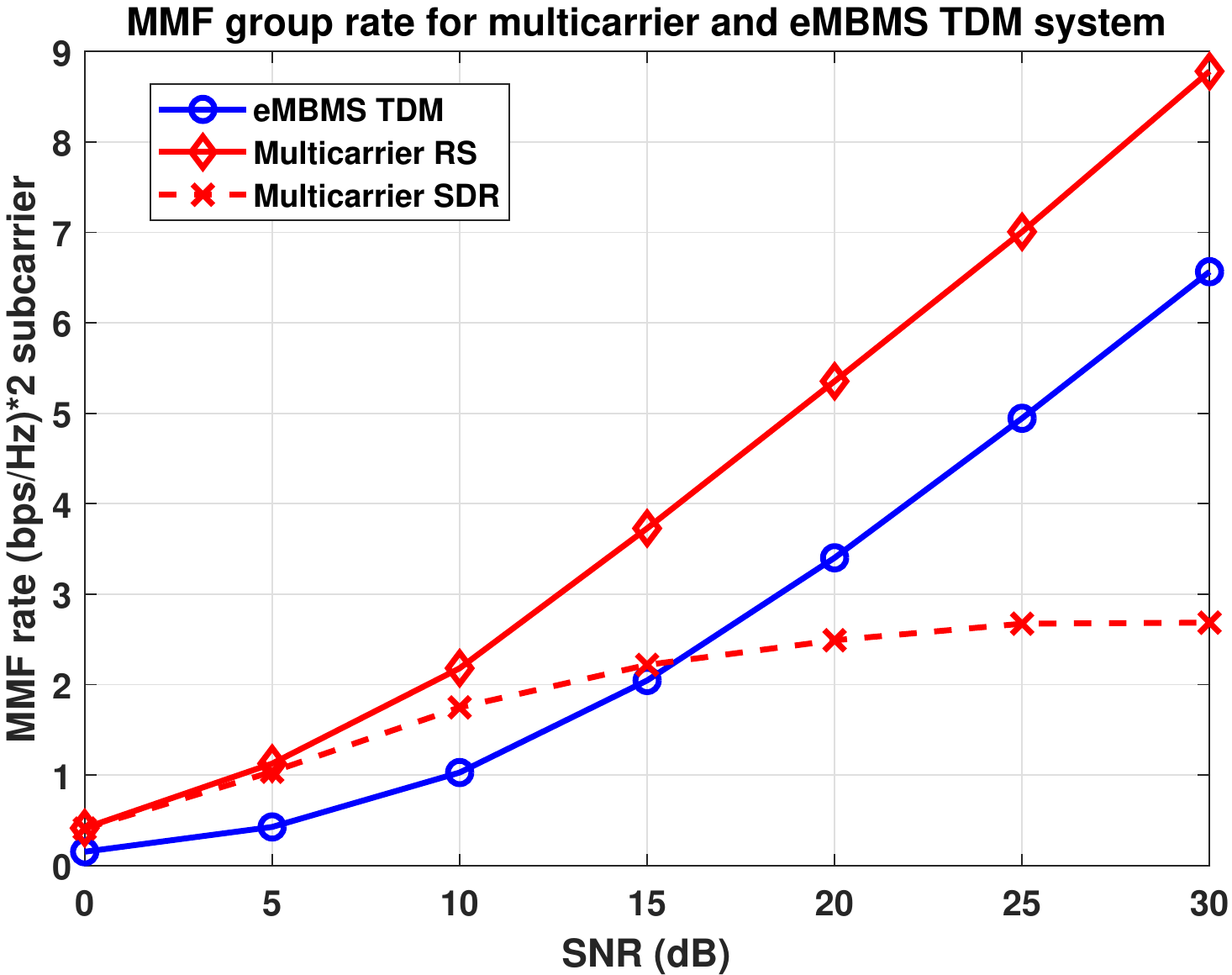}
	\caption{Corresponding MMF rate for 2-2-2-2 system}
	\label{fig:fig1}
\end{figure}

In the first simulation scenario, we consider a 2-2-2-2 multicarrier system i.e. Available Subcarriers-Transmit  Antennas-Multicast group-Users per group and every two users required the same multicast message. In this situation, each OBB program is separated into two different elements and these two elements are transmitted on the two subcarriers respectively. This model reaches the transmitter limit i.e. $2\leq1+(2-1)*2=3$ and it achieves the highest and lowest DoF simultaneously on both subcarriers when using RS i.e. 
$\text{D}_{sys}=\frac{1}{1+(M-1)}=\frac{1}{2}$. TDM system transmits the multiple services in a interference-free manner but on multiple time slots, where inside one time slot, different elements are assigned to different subcarriers as described in the system mode. Note that in the result shown in Fig.~\ref{fig:fig1}, the Y-axis is the sum MMF-rate for two subcarriers since the multicarrier system is jointly optimized over two subcarriers. From the figure we can see that, the classic SDR-based beamforming (shown as Multicarrier SDR), gives the upper bound of the sum MMF-rate which outperforms the current TDM system at low SNR region (0 to 15dB). Due to the performance saturation caused by inter-group interference, TDM system provides better sum MMF-rate at higher SNR region; With RS, the saturation is avoided and the gains of RS over the classic beamforming as well as TDM are very pronounced.%Results in Fig.1 show that the RS-based approach significantly outperforms other two methods.

\begin{figure}[t]
	\centering
	\includegraphics[angle=0,width=0.45\textwidth]{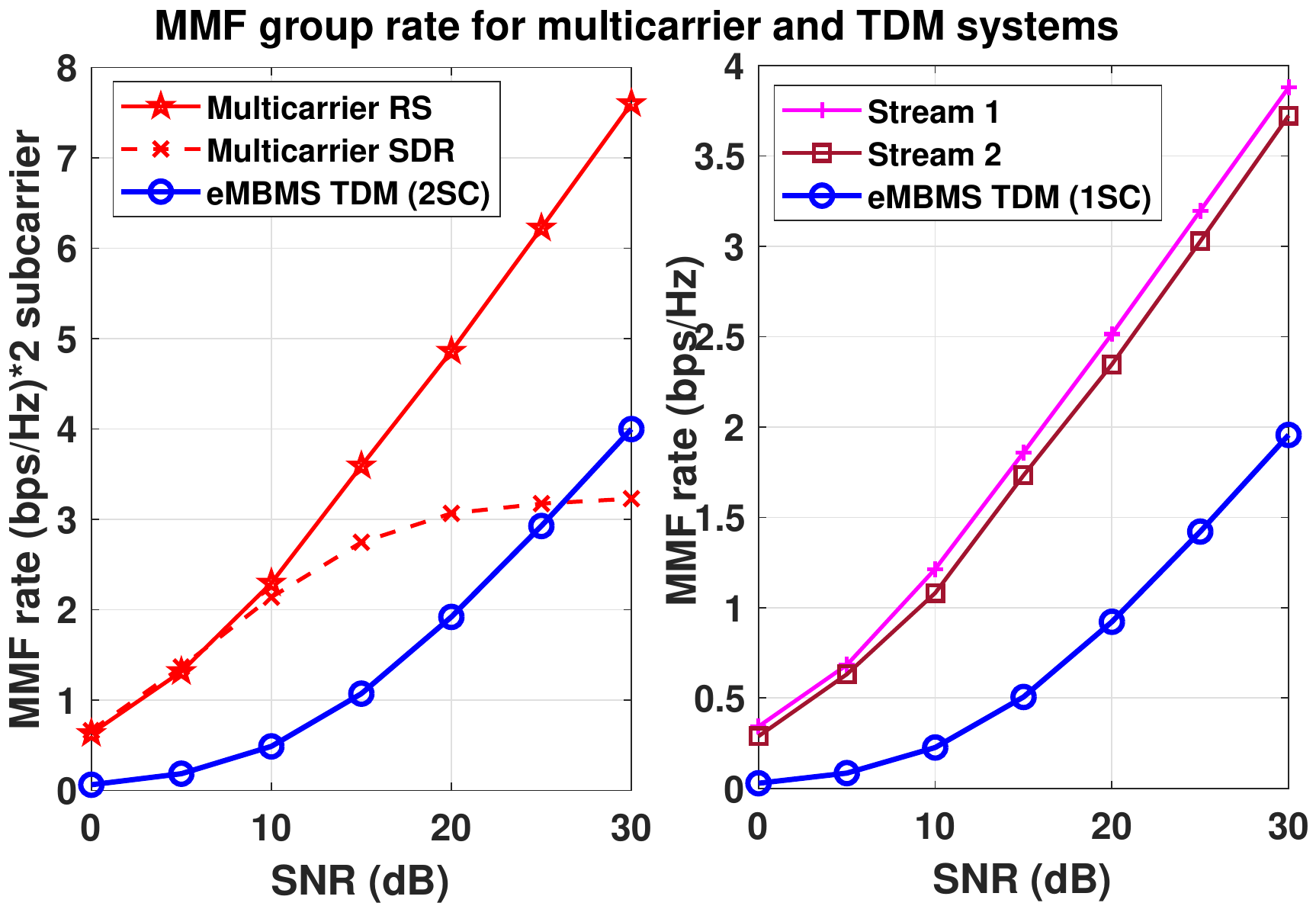}
	\caption{Corresponding MMF rate for 2-4-3-3 system}
	\label{fig:fig2}
\end{figure}

In the second simulation, we test a 2-4-3-3 system, where the corresponding user configuration is set in a similar manner as previous but with 3 users per group. This model sets 2 multicast groups as degraded groups and achieves $\text{D}_{sys}=\frac{1}{3}$. The result shown in Fig.~\ref{fig:fig2} illustrates that multigroup multicast integrates more group message into one subcarrier drives the MMF rate difference between RS and TDM even larger (from around 2.2bps/Hz to 3.5bps/Hz @30dB SNR). Stream 1 and 2 in the right-hand side figure represents the MMF-rate for subcarrier 1 and 2. Comparing two sub-figures in Fig.~\ref{fig:fig2}, it shows that the multicarrier rate is equivalent to the summation of single carrier rates for each service stream, which shows that the RS-based approach not only provides the non-saturated higher MMF rate than TDM, but also achieves the fairness between all the subcarriers in terms of the MMF rate.

\section{Conclusion}
This paper considered the optimization problem of RS-based transmitter beamforming in typical multicarrier multigroup multicasting scenarios. We illustrated that the RS-based approach not only provides higher MMF rate/spectrum efficiency compared to the current TDM-based multicasting within the same time and frequency resource, but also exhibits strictly higher and non-saturated MMF-rates compared to the classic SDR-based beamforming method. We also showed the potential of applying RS in a multicarrier scenario which is close to the systems in practice, e.g., eMBMS in LTE. The considered system can be further developed to be closer to the practical scenario by assuming the imperfect channel estimation, also introducing the RS into the typical transceiver chain and generating bit/block error rate curve can also be of interest in the future work.






\bibliographystyle{IEEEtran}
\bibliography{reference}

\end{document}